\documentclass[preprint,aps]{revtex4}

\usepackage{graphicx}
\usepackage{dcolumn}
\usepackage{bm}
\usepackage{epstopdf}

\begin{document}
\title{Broadband energy squeezing and tunneling based on unidirectional modes}

\author{Lujun Hong$^{1}$, Yazhou Wang$^2$, Yun Shen$^{1,6}$, Xiaohua Deng $^{1}$, Kai Yuan$^{3}$, Sanshui Xiao$^2$, and Jie Xu $^{4,5,7}$}
\address{$^1$ Institute of Space Science and Technology, Nanchang University, Nanchang 330031, China}
\address{$^2$ DTU Fotonik, Department of Photonics Engineering, Technical University of Denmark, DK-2800 Kgs. Lyngby, Denmark}
\address{$^3$ Information Engineering School of Nanchang University, Nanchang University, Nanchang 330031, China}
\address{$^4$ School of Medical Information and Engineering, Southwest Medical University, Luzhou 646000, China}
\address{$^5$ Medicine \& Engineering \& Information Fusion and Transformation Key Laboratory of Luzhou City, Luzhou 646000, China}
\address{$^6$ shenyun@ncu.edu.cn}
\address{$^7$ xujie011451@163.com}



\begin{abstract}
Energy squeezing attracts many attentions for its potential applications in electromagnetic (EM) energy harvesting and optical communication. However, due to the Fabry-Perot resonance, only the EM waves with discrete frequencies can be squeezed and, as far as we know, in the previous energy-squeezing devices, stringent requirements of the materials or the geometrical shape are needed. We note that the structures filled with epsilon-near-zero (ENZ) mediums as reported in some works can squeeze and tunnel EM waves at frequencies (e.g. plasma frequency). However, the group velocity is usually near zero which means few EM information travel through the structures. In this paper, low-loss energy squeezing and tunneling (EST) based on unidirectional modes were demonstrated in YIG-based one-way waveguides at microwave frequencies. According to our theoretical analysis and the simulations using finite element method, broadband EST was achieved and the EM EST was observed even for extremely bended structures. Besides, similar EM EST was achieved in a realistic three-dimensional remanence-based one-way waveguide as well. The unidirectional modes-based EST paving the way to ultra-subwavelength EM focusing, enhanced nonlinear optics, and designing numerous functional devices in integrated optical circuits such as phase modulator.

\end{abstract}

\maketitle

\section{Introduction}

One-way Electromagnetic (EM) waves propagating in systems in only one direction are similar to those one-way edge modes founded in the quantum Hall effect\cite{Prang:Th}. The one-way EM modes was experimentally observed in 2009 by using a microwave magneto-optical (MO) material, i.e. yttriun-iron-garnet (YIG)\cite{Wang_2009}. This kind of one-way EM modes sustained at the MO interface are named surface magnetoplasmon (SMP) and utilizing an external magnetic field to break the time-reversal symmetry is a fundamental way to induce such one-way SMP\cite{Brion_1972,Wallis:Th}. Recently, Tsakmakidis and his colleagues proposed an one-way waveguide consisting of one layer of semiconductor (InSb) and in the waveguide, one metal layer was set to stop the EM wave, and it was reported that the time-bandwidth limit in such one-way waveguide can be broken\cite{Tsakmakidis_2017}. More recently, we proposed several similar one-way waveguides and interesting phenomenons such as slowing wave and truly rainbow trapping were reported as well\cite{Xu:Br,Xu:Sl}.

Squeezing electromagnetic waves at subwavelength or even deep subwavelength scale is important for near-field microscopy\cite{Atkin:Na,Berweger:Mi}, semiconductor laser\cite{Machida:Ob}, optical sensing\cite{Frascella:Ov} and nonlinear optics\cite{Tucker:Ph,Thapliyal:Hi}. Epsilon-near-zero (ENZ) materials refers to those materials with a near-zero relative permittivity and owing to the special physical characters, they are widely used in researches of metamaterial and metadevices. In the past two decades, researchers reported several types of devices based on ENZ materials, in which the EM energy squeezing and tunneling (EST) were observed\cite{silveirinha_2006,liu_2008,liu_2009,Jing_2010,li_2017}. Moreover, utilizing metamaterials and metasurface are believed to be an efficient way to realize EM EST\cite{xu_2018,Liang:Sq}. Recently, it was reported that, by utilizing two-dimensional materials (graphene and hexagonal boron nitride), an ultimate confinement limit of the length scale of only one atom was achieved based on a graphene-insulator-metal structure\cite{iranzo_2018}.

However, the EM EST always suffers from the Fabry-Perot resonance and reflections, and the coupling efficiency of the guided modes is always low. Therefore, to achieve EM EST, researchers are required to carefully optimize the geometric shape and sift the materials. On the other hand, one-way waveguide can support unidirectionally propagating EM waves without backscattering and those one-way EM waves, as reported in many works, can nearly perfectly bypass disorders or bends. In this paper, we theoretically study the propagation characters in metal-dielectric-YIG-metal (MDYM) structures and by utilizing COMSOL software, EM EST are achieved and verified in the YIG-based one-way waveguides. Moreover, the EM EST in such waveguides are proved to be low-loss. Furthermore, similar EM EST are achieved in a remanence-based three-dimensional (3D) MDYM waveguide as well. This kind of unidirectional modes-based EM EST is useful for the oncoming optical communication system.

\section{Broadband Energy Squeezing based on unidirectional modes}

As reported in many works, the effect of energy squeezing is usually achieved in cavities surrounded by metallic materials. Here, we propose a novel waveguide performed like a energy compressor based on unidirectional EM modes. The inset of Fig. 1(a) shows the schematic of that waveguide, which is composed of two layers of metal layers (treated as perfect electric conductor(PEC) in microwave regime), one layer of dielectric (air) and one layer of YIG under an external magnetic field $H_0$, i.e. the MDYM structure. The asymmetric relative permeability tensor of YIG in this case takes the following form
\begin{equation}
	\bar{\mu}_\mathrm{h}=\left[\begin{array}{ccc}
		\mu_\mathrm{r} & -i \mu_\mathrm{k} & 0 \\
		i \mu_\mathrm{k} & \mu_\mathrm{r} & 0 \\
		0 & 0 & 1
	\end{array}\right]
\end{equation}
with ${\mu _\mathrm{r}} = 1 + \frac{{\omega
_\mathrm{m}}{(\omega _0-i\alpha{\omega})}}{{{(\omega _0-i\alpha{\omega})}^2-{\omega}^2}}$ and ${\mu _\mathrm{k}} = \frac{{{\omega _\mathrm{m}}{\omega}}}{{{(\omega _0-i\alpha{\omega})}^2} - {\omega}^2}$. ${\omega}$, ${\alpha}$, ${\omega _\mathrm{m}}$, ${\omega _0} =2{\pi}{\gamma}{\mathrm{H}_0}$ (${\gamma}=2.8\times10^6$ rad/(sG), gyromagnetic ratio) are the angular frequency, the damping coefficient, the characteristic circular frequency and the precession angular frequency, respectively\cite{Deng_2015}. In the MDYM model, only the transverse electric (TE) modes ($k_\mathrm{z}=0$) are sustained on the air-YIG interface, and deriving from the Maxwells' equations and two boundary conditions in y direction, the dispersion relation of the TE modes (SMPs) can be written as below
\begin{eqnarray}
{\alpha _\mathrm{d}}{\mu _v}+[\frac{\alpha _\mathrm{h}}{\tanh({\alpha_\mathrm{h}} h)}+\frac{\mu_\mathrm{k}}{\mu_\mathrm{r}}k]\tanh({\alpha _\mathrm{d}} d)=0
\end{eqnarray}
where ${\alpha _\mathrm{d}} = \sqrt{{k}^2 - {\varepsilon _\mathrm{d}}{k _0}^2}$ (${k _0=\omega/c}$) and ${\alpha _\mathrm{h}} = \sqrt{{k}^2 - {\varepsilon _\mathrm{h}}{\mu_v}{k _0}^2}$ (${\mu_v}={\mu_\mathrm{r}}-{\mu_\mathrm{k}}^2/{\mu_\mathrm{r}}$, the Voigt permeability) demonstrate the attenuation coefficients of the SMPs in the air and YIG layers\cite{Shen_2019}. In Eq. (2), by letting $k\to +\infty$ and $k\to -\infty$, we calculated two asymptotic frequencies of the SMPs, i.e. $\omega_\mathrm{AF}^+=\omega_0+\omega_\mathrm{m}$ and $\omega_\mathrm{AF}^-=\omega_0+\omega_\mathrm{m}/2$. As an example, the dispersion diagram of the MDYM is shown in Fig. 1(a) for $d=0.1\lambda_\mathrm{m}$ ($\lambda_\mathrm{m}=2\pi c/\omega_\mathrm{m}$) and $h=0.15\lambda_\mathrm{m}$. The red line indicates the dispersion curves of SMPs and there is a clear one-way propagation (OWP) band (shaded rectangle area) limited by two AFs. Moreover, the cyan shaded zones and the dot-dashed line respectively represent the bulk zones in the YIG layer and the light cone of air, respectively. Different with the SMPs in infinite YIG based one-way waveguide, the SMPs in the MDYM waveguide strongly coupled with bulk modes when the frequency is near $\omega_\mathrm{AF}^+$, making the horizontal dispersion curve branch in the $k>0$ region. To explain the theory of energy squeezing in MDYM structures, in Fig. 1(b), the zoomed-in dispersion curves of SMPs are plotted for three different values of $d$, i.e. $d=0.1\lambda_\mathrm{m}$ (red line), $d=0.03\lambda_\mathrm{m}$ (black line) and $d=0.01\lambda_\mathrm{m}$ (cyan line). As a result, when reducing the thickness of the air layer, the dispersion curves rose up and the dispersion branch of $\omega<\omega_\mathrm{AF}^-$ gradually disappeared, in another word, the dispersion valley in thick air case ($d=0.1$) gradually rose up and in the thin air case ($d=0.01\lambda_\mathrm{m}$), no SMPs are found in the frequency band $\omega<\omega_\mathrm{AF}^-$. Most importantly, although the dispersion rose up or 
dropped down when changing the values of $d$, the whole OWP band remain the same, which implies that the EM modes with frequencies falling within the OWP band can always propagate to only one direction even in ultra-thin or ultra-thick air-based MDYM structure. Therefore, as shown the inset of Fig. 1(b), we believe that the one-way propagating EM waves in the left thicker part can always nealy perfectly couple with the sustained EM modes in the right part of the waveguide for the one-way propagation character in both parts.

\begin{figure}[t!]
\centering\includegraphics[width=6 in]{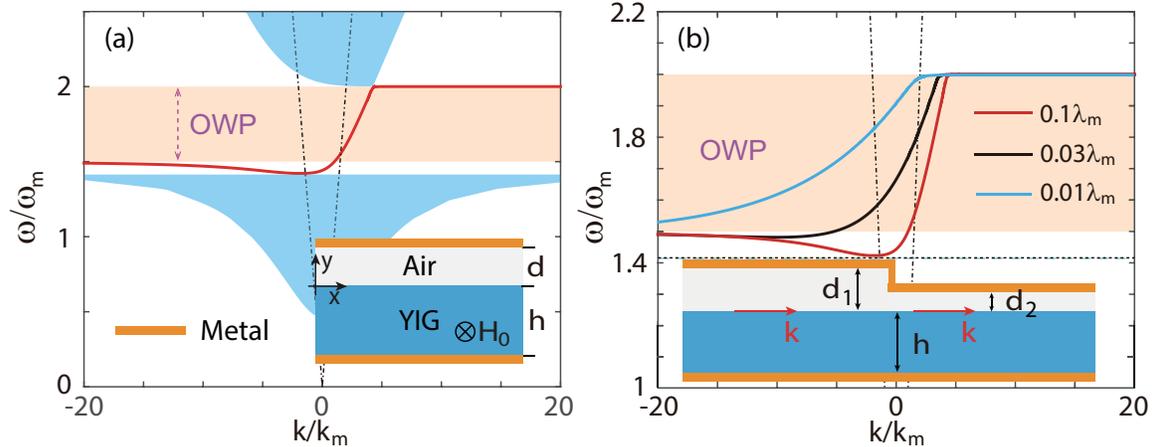}
\caption{(Color online) (a) Dispersion diagram of the MDYM configuration as  $h=0.15\lambda_\mathrm{m}$ and $d=0.1\lambda_\mathrm{m}$. (b) Dispersion curves of SMPs for three different $d$, i.e. $d=0.1\lambda_\mathrm{m}$ (red line), $d=0.03\lambda_\mathrm{m}$ (black line) and $d=0.01\lambda_\mathrm{m}$ (cyan line), as $h=0.15\lambda_\mathrm{m}$. The inset of (a) shows a straight MDYM waveguide while the inset of (b) demonstrates a EM compressor based on MDYM configuration. The other parameters are $\omega_0=\omega_\mathrm{m}$,  $\lambda_\mathrm{m} \approx 60$ mm, $\varepsilon_\mathrm{d}=1$, $\varepsilon _\mathrm{h}=15$, $\alpha=0$ and $\mathrm{H}_0 \approx 1785$ G.}\label{fig1}
\end{figure}
\begin{figure}[b!]
	\centering\includegraphics[width=5 in]{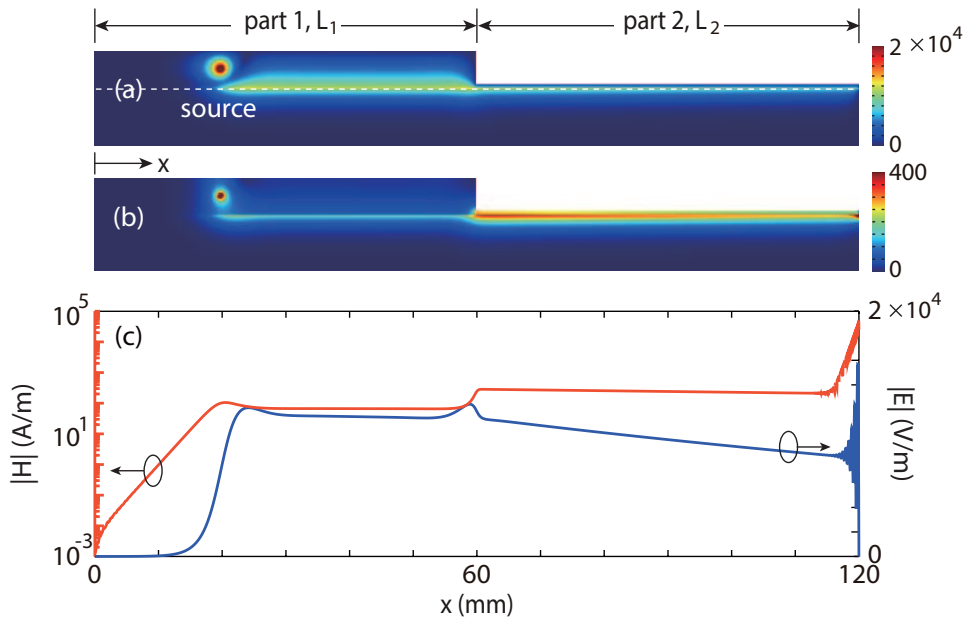}
	\caption{(Color online) Simulated (a) electric field and (b) magnetic field distributions as $d_1=0.1\lambda_\mathrm{m}$, $d_2=0.1\lambda_\mathrm{m}$ and $h=0.15\lambda_\mathrm{m}$. (c) The amplitudes of the electric field and magnetic field along the air-YIG interface. $L_1=L_2=60$ mm and the source was located at $x=L_1/3$.}\label{fig2}
\end{figure}

\begin{figure}[t!]
	\centering\includegraphics[width=6 in]{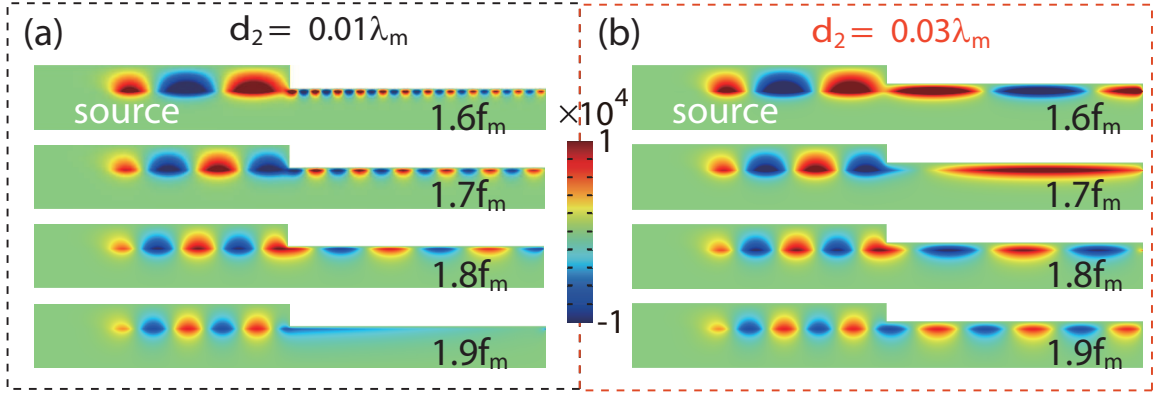}
	\caption{(Color online) Simulations for multi-frequencies one-way propagating EM waves for (a) $d_2=0.01\lambda_\mathrm{m}$ and (b) $d_2=0.03\lambda_\mathrm{m}$. Four working frequencies from top to bottom are respectively $1.6f_\mathrm{m}$, $1.7f_\mathrm{m}$, $1.8f_\mathrm{m}$ and $1.9f_\mathrm{m}$. The other parameters are the same as in Fig. 2.}\label{fig3}
\end{figure}

\begin{figure}[t!]
	\centering\includegraphics[width=5.5 in]{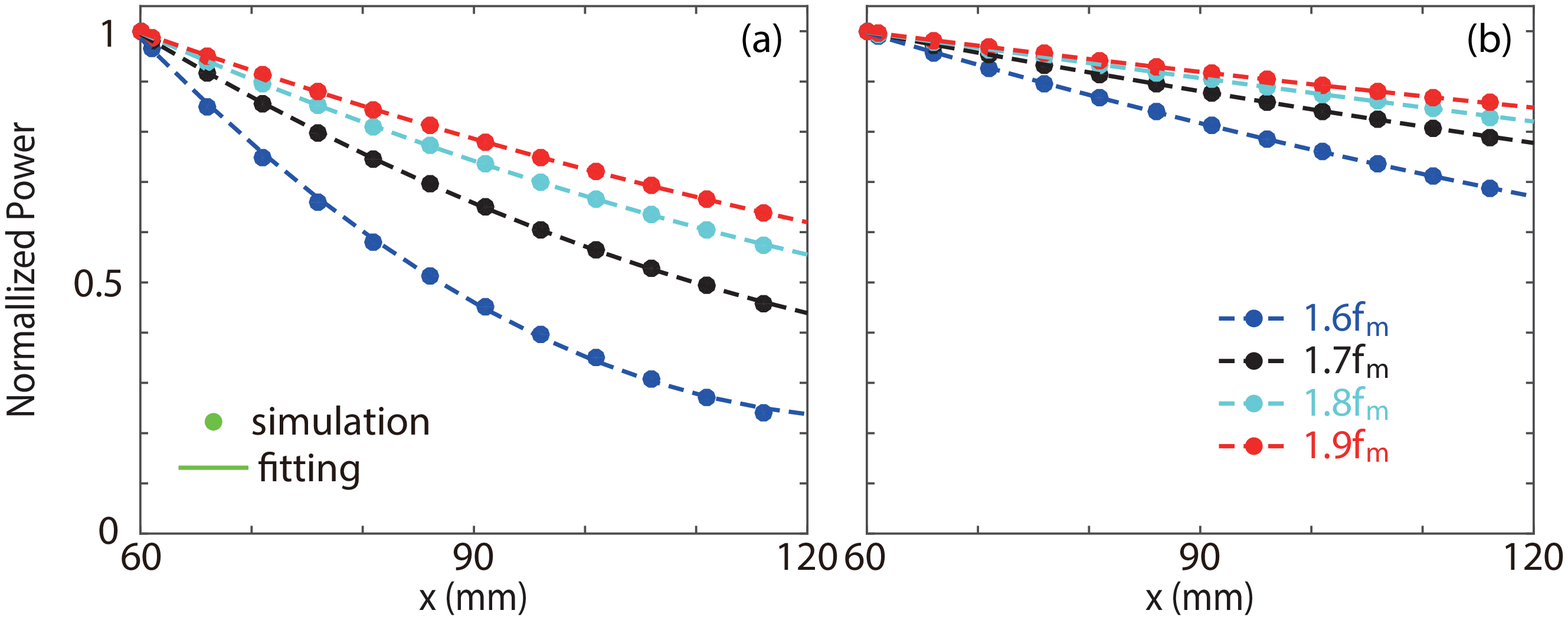}
	\caption{(Color online) Analysis of the transmission efficiencies in simulations shown in (a) Fig. 3(a) and (b) Fig. 3(b). }\label{fig4}
\end{figure}

Note that the loss of MO materials, e.g. YIG, has no significant impact on the one-way propagation character of the SMPs in one-way waveguides\cite{Tsakmakidis_2017,Xu:Br}. In this paper, the loss effects are considered by setting $\alpha=0.001$ in the simulations using finite element method (FEM). In Fig. 2, we performed a full wave simulation to reveal the high performance energy squeezing based on our proposed MDYM configuration. Figs. 2(a) and 2(b) are the simulated electric field and magnetic field distributions, respectively. One can clearly see that the excited EM wave unidirectionally propagate to the end of part 1 of the waveguide and further propagate to the narrow part (part 2) until it reaches to right most end. In addition, no interference was observed in both Figs. 2(a) and 2(b), which is in agreement with above analysis. More clear results of such supercoupling between the EM modes in part 1 and part 2 are shown in Fig. 2(c), demonstrating the electric field (blue line) and the magnetic field (red line) distributions along the air-YIG interface. Different with the decreased amplitude of the electric field in part 2, the magnetic field is significantly enhanced in the narrow region and the magnetic field reaches the maximum value on the right terminal end which is set to be a PEC wall. Due to the ability of immune to back-scattering, this kind of energy squeezing should be high performance and here we emphasis that, compared to the SPP- or ENZ-based structures,  our proposed thick-thin MDYM waveguides can squeeze SMPs with frequencies falling within a continuous broadband frequency band, i.e. the OWP band.

In Fig. 3, several FEM simulations were performed to show the broadband squeezing property of our proposed thick-thin MDYM waveguide and in this case we focused on the distributions of the electric field component ($E_\mathrm{z}$). Four working frequencies, i.e. $1.6f_\mathrm{m}$, $1.7f_\mathrm{m}$, $1.8f_\mathrm{m}$ and $1.9f_\mathrm{m}$ were chose in the simulations. As a result, complete one-way propagating EM waves and clear energy squeezing were observed in the eight simulations in Fig. 3. Moreover, all of the excited EM waves propagates to $+x$ direction and, in part 1, according to the dispersion relation shown in Fig. 1(b) (cyan line), the higher the frequency, the larger the values of $|k|$ (the wavenumber), which makes the phase of $E_\mathrm{z}$ reaches $2\pi$ easier for the case of $f=1.9f_\mathrm{m}$ than the case of $f=1.6f_\mathrm{m}$. In part 2, the phase changing become complicate since the values of $k$ can be negative or positive. In a word, Fig. 3(a) indicates that our designed MDYM structure has the ability of squeezing EM waves with frequencies in a broadband OWP region (in this case the bandwidth of the OWP band equals $0.5\omega_\mathrm{m}$). On the other hand, as shown in Fig. 3(b), the thickness of air in part 2 of the MDYM waveguide is three times of the one in simulations shown in Fig. 3(a), and the EM waves were squeezed in both cases, which proves that changing the vertical thickness of the air cannot break the process of energy squeezing in the MDYM structure. The transmission efficiencies in +x direction of the FEM simulation in Fig. 3 are illustrated in Fig. 4 for $L_1\leq x \leq L_1+L_2$. It is obvious that, for both cases, i.e. $d_2=0.01\lambda_\mathrm{m}$ case and $d_2=0.03\lambda_\mathrm{m}$ case, the higher the frequencies of the propagating EM waves, the lesser the propagation loss. In addition, the loss in the case of thicker air ($d_2=0.03$) is much lesser than the one in the case of $d_2=0.01$ since more energy decayed in the YIG layer. It is worth to note that even the losses seem like huge and only near $20\%$ energy remains for the case of $f=1.6f_\mathrm{m}$ as shown in Fig. 4(a), one should notice that if we set $L_2=10$ mm, more than $80\%$ energy will remain and the corresponding transmission efficiency can be more than $93\%$ when considering $f=1.9f_\mathrm{m}$. Therefore, the thick-thin MDYM structure can perfectly squeeze unidirectionally propagating EM waves with high efficiency regardless of the thickness of the air at subwavelength scale.

\section{Robust energy squeezing and tunneling}

\begin{figure}[t!]
\centering\includegraphics[width=5.5 in]{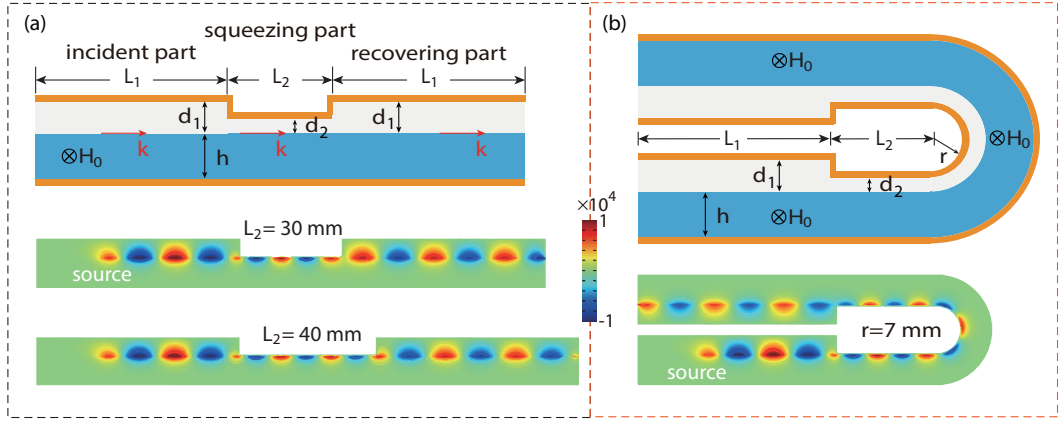}
\caption{(Color online) Two kinds of U-type MDYM channel. The schematic and simulated $E_\mathrm{z}$ for (a) straight channel and (b) bended channel with $L_2=30$ mm and $r=7$ mm. For both (a) and (b), the working frequency was set to be $f=1.75f_\mathrm{m}$. The other parameters are the same as in Fig. 3(a).}\label{fig5}
\end{figure}
One of the most important advantage of one-way waveguide is the robustness of the unidirectional propagation character of SMPs, which in our opinion make it possible to squeeze and recover the waves in a U type MDYM channel. In Fig. 5, we designed two kinds of U-type channel and the first channel is straight with the parameters are the same as the one in Fig. 3(a) except for the values of $L_2$, while the second channel is bended with the radius of the corner $r=7$ mm. For the first channel, we investigate the energy squeezing and tunneling (EST) in cases of different $L_2$. According to the simulation results illustrated in Fig. 5(a), in both simulations the excited waves unidirectionally propagated to the right most end of the waveguide and were completely squeezed in the squeezing part. More importantly, as we expect, the EM waves recovered in the recovering part with no obvious reflection. Another striking result shown in Fig. 5(a) is that the length of the squeezing part only influence the phase of the EM waves at the end surface of the squeezing part and it has no obvious impact on the supercoupling between the SMPs in the thick and the thin MDYM waveguide. Since the supercoupling effects remain the same regardless the values of the thickness of the air and the length of the squeezing part, our proposed unidirectional SMPs-based EST are untouched by the Fabry-Perot resonance. In Fig. 5(b), we designed a extraordinarily bended U-type MDYM channel and the simulated electric field component $E_\mathrm{z}$ is shown in the second diagram. According to the simulation results, the excited unidirectional EM wave was squeezed in the squeezing part and channelled through the half-circle shape bend, and recovered in the recovering part of the MDYM structure. Again the loss effects can be greatly eliminated by carefully designing the waveguide parameters $L_1$ and $r$. 

Besides, the dispersion characters of the SMPs in the MDYM structures can be engineered by controlling the external magnetic field, which will further correction the OWP band. Therefore, as shown in Fig. 6, two AFs (red and black lines) are plotted as functions of $\omega_0$ ($\omega_0=2\pi \gamma H_0$) and the colored zone indicated the OWP regions for $0<\omega_0\leq \omega_\mathrm{m}$. As a result, according to our calculation, the robust EST can be achieved in an appreciable band  ($0.5\omega_\mathrm{m}<\omega<2\omega_\mathrm{m}$) by optimizing the external magnetic field. In order to verify the exactness of the conclusion, we performed FEM simulations in the straight U-type MDYM structure shown in Fig. 5(a) with $\omega_0=0.5\omega_\mathrm{m}$. Two working frequencies are chosen, i.e. $f=1.2f_\mathrm{m}$ (marked by the lower green star) and $f=1.4f_\mathrm{m}$ (marked by the upper green star) and for both frequencies, the EM waves were squeezed and recorvered just like above simulations shown in Fig. 5. Consequently, owing to the capability of unidirectionally squeezing and recovering EM energy, the MDYM configuration (or one-way waveguide) is proved to be a powerful and high performance candidate in optical communication and optical integrated circuit.

\begin{figure}[t!]
\centering\includegraphics[width=3.5 in]{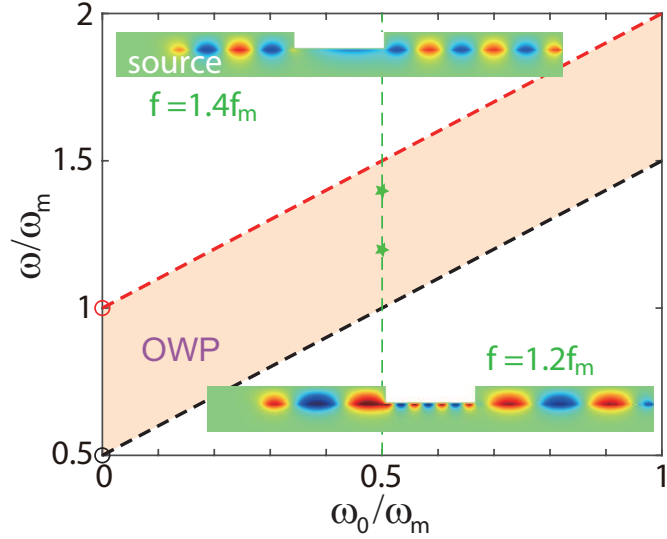}
\caption{(Color online) The AFs (or the OWP band) as a function of $\omega_0$. The insets indicate the simulated electric field distribution for $f=1.4f_\mathrm{m}$ (upper inset) and $f=1.2f_\mathrm{m}$ (lower inset) when $\omega_0=0.5\omega_\mathrm{m}$ (green dashed line) which corresponds to $H_0\approx892.5$ G.}\label{fig6}
\end{figure}

\section{A 3D electromagnetic compressor based on remanence}
\begin{figure}[t!]
	\centering\includegraphics[width=5 in]{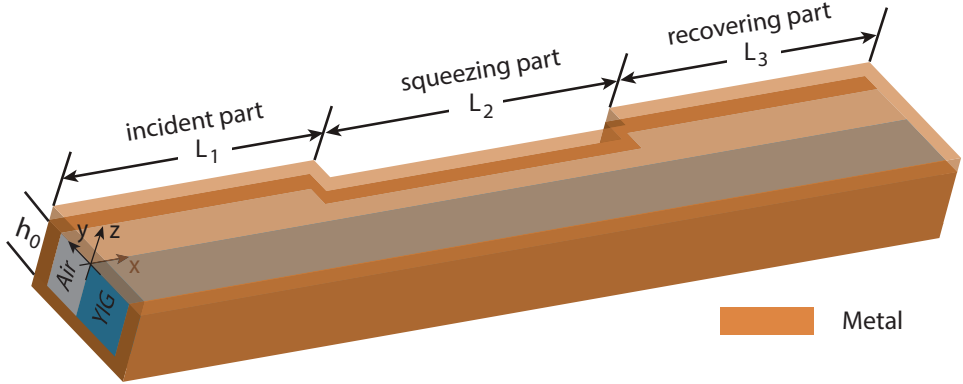}
	\caption{(Color online) Schematic of a 3D MDYM channel based on remanence. }\label{fig7}
\end{figure}

\begin{figure}[t!]
	\centering\includegraphics[width=5 in]{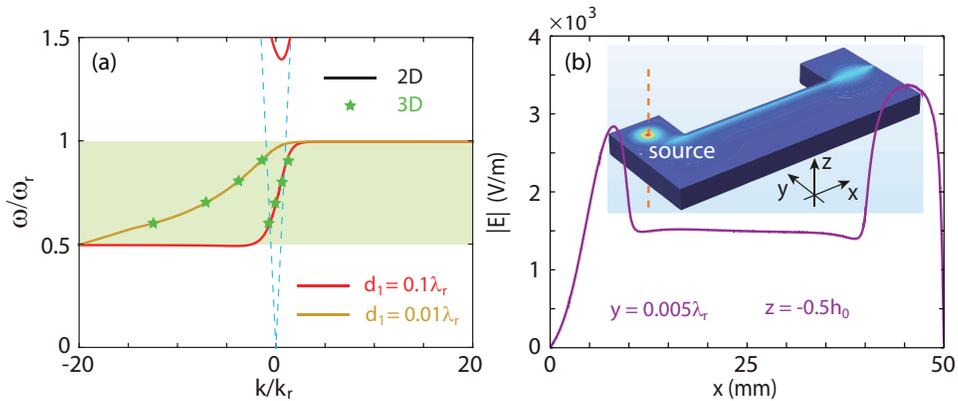}
	\caption{(Color online) (a) Dispersion diagram of 2D (solid lines) and 3D (marked by stars) MDYM structures for $d_1=0.1\lambda_\mathrm{r}$ (red lines) and for $d_1=0.01\lambda_\mathrm{r}$ (yellow line). (b) Simulated electric field distribution in the 3D simulation as $f=0.75f_\mathrm{r}$ along the intersection line of a XZ plane ($y=0.005\lambda_\mathrm{r}$) and a XY plane ($z=-0.5h_0$). The other parameters are $\omega_\mathrm{r}=2\pi\times3.587\times10^9$ rad/s, $d_1=0.1\lambda_\mathrm{r}$, $d_2=0.01\lambda_\mathrm{r}$, $h_0=4$ mm and $\alpha=0.001$. }\label{fig8}
\end{figure}

YIG-based one-way waveguides, as reported in many works, sustain the transverse electric (TE) modes rather transverse magnetic (TM) modes, making it possible to cover the structure in the direction of $H_0$ with no influence to the one-way propagation character of the SMPs\cite{Xu:Ul}. Considering the realistic 3-dimensional (3D) condition, in this subsection, we design the 3D MDYM channel based on remanence. Fig. 7 demonstrates the schematic of our designed 3D channel, in which the air and YIG are surrounded by four layers of metal and the channel is divided into three parts, i.e. the incident part, the squeezing part and the recovering part. The lengths of three parts are respectively $L_1$, $L_2$ and $L_3$ while the height of the channel ($h_0$) was set to be 4 mm in this paper. 

The relative permeability of YIG in this subsection takes the form\cite{Pozar:Mi}
\begin{equation}
	\bar{\mu}=\left[\begin{array}{ccc}
		\mu_1 & -i \mu_2 & 0 \\
		i \mu_2 & \mu_1 & 0 \\
		0 & 0 & 1
	\end{array}\right]
\end{equation}
where $\mu_1=1+\frac{i\alpha\omega_\mathrm{r}}{\omega(1+\alpha^2)}$ and $\mu_2=-\frac{\omega_\mathrm{r}}{\omega(1+\alpha^2)}$. $\omega_\mathrm{r}=2\pi\times3.587\times10^9$ rad/s represents the characteristic circular frequency. The dispersion relation in the remanence-based 2D MDYM structure can directly derived from Eq. (2) by replacing the attenuation efficient $\alpha_\mathrm{h}$ by $\alpha_\mathrm{h1}=\sqrt{k^2-\varepsilon_\mathrm{h} \mu_\mathrm{v1} k_0^2}$ ($\mu_\mathrm{v1}=\mu_1-\mu_2^2/\mu_1$). As shown in Fig. 8(a), the solid lines indicate the dispersion curves of SMPs for $d=0.1\lambda_\mathrm{r}$ (red line) and $d=0.01\lambda_\mathrm{r}$ (yellow line) and $\lambda_\mathrm{r}=2\pi \mathrm{c}/\omega_\mathrm{r}\approx 83.6$ mm. The stars in Fig. 8(a) represent the corresponding solutions of the 3D dispersion equation by using the mode analysis module of COMSOL and we found that all the stars is in the solid lines. In the 3D simulation, we set $\alpha=0.001$, $L_1=L_3=10$ mm, $L_2=30$ mm and $h=0.15\lambda_\mathrm{r}$. The thicknesses of the air of the incident, squeezing and recovering part are respectively $d_1=0.1\lambda_\mathrm{r}$, $d_2=0.01\lambda_\mathrm{r}$ and $d_3=0.1\lambda_\mathrm{r}$. The inset of Fig. 8(b) shows the distribution of the simulated electric field and we found that, similar with above 2D simulations, the EM wave excited by the source at $x=L_1/2$ and it channel through the squeezing part and transmit into the recovering part and the energy accumulating at the end surface. The purple line indicates the amplitudes of the electric field along the intersection line of $y=d_2/2$ and $z=h_0/2$. Moreover, the smooth line also demonstrates that the EM wave travel through the waveguide without reflection or backscattering and the EM energy is truly squeezed and recovered in such a remanence-based MDYM channel.

\section{Conclusion}
In conclusion, we have proposed a novel metal-dielectric-YIG-metal (MDYM) configuration and theoretically investigated the capability of such MDYM structure in energy squeezing and tunneling (EST). Based on the theoretical analysis and the simulations using COMSOL software, the MDYM configuration, owing to the one-way propagation character of the SMPs, has been proved to be high-performance in squeezing the EM energy and the corresponding transmission efficiencies can be more than $93\%$ when the parameters of the waveguides are carefully engineered. Besides, continuous and broadband EST were achieved based on the one-way SMPs in MDYM structures, which, to our knowledge, has never been reported before. The vertical and lateral lengths of the squeezing part has also be discussed and, interestingly, different with the energy squeezing based on ENZ materials, Fabry-Perot resonance has no obvious impact on the capability of energy squeezing in such MDYM waveguides. Moreover, in two kinds of our designed U-type MDYM channel, due to the supercoupling between the one-way EM modes in incident, squeezing amd recovering parts, the EM waves traveled unidirectionally and channelled through the U-type MDYM channel even when the squeezing part is greatly bended. Finally, we designed a realistic 3D MDYM channel for EST based on remanence and similar EST were observed in such 3D structure. Our findings pave a way to squeeze energy at ultra-subwavelength scale and the theory proposed in this paper is highly possible to be applied in terahertz or even visible regime.

\section*{Fundings}

National Natural Science Foundation of China (NSFC) (61927813, 61865009, 11904152); Start-up funding of Southwest Medical University (20/00040186); Department of Science and Technology of Sichuan Province (14JC0153); Science and Technology Strategic Cooperation Programs of Luzhou Municipal People’s Government and Southwest Medical University (2019LZXNYDJ18).

\newpage
\bibliography{squeezing}

\begin{thebibliography}{25}
\expandafter\ifx\csname natexlab\endcsname\relax\def\natexlab#1{#1}\fi
\expandafter\ifx\csname bibnamefont\endcsname\relax
  \def\bibnamefont#1{#1}\fi
\expandafter\ifx\csname bibfnamefont\endcsname\relax
  \def\bibfnamefont#1{#1}\fi
\expandafter\ifx\csname citenamefont\endcsname\relax
  \def\citenamefont#1{#1}\fi
\expandafter\ifx\csname url\endcsname\relax
  \def\url#1{\texttt{#1}}\fi
\expandafter\ifx\csname urlprefix\endcsname\relax\def\urlprefix{URL }\fi
\providecommand{\bibinfo}[2]{#2}
\providecommand{\eprint}[2][]{\url{#2}}

\bibitem[{\citenamefont{Prange and Girvin}(1987)}]{Prang:Th}
\bibinfo{author}{\bibfnamefont{R.~E.} \bibnamefont{Prange}} \bibnamefont{and}
  \bibinfo{author}{\bibfnamefont{S.~M.} \bibnamefont{Girvin}},
  \emph{\bibinfo{title}{The Quantum Hall Effect}}
  (\bibinfo{publisher}{{Springer}}, \bibinfo{year}{1987}).

\bibitem[{\citenamefont{Wang et~al.}(2009)\citenamefont{Wang, Chong,
  Joannopoulos, and Solja{\v{c}}i{\'{c}}}}]{Wang_2009}
\bibinfo{author}{\bibfnamefont{Z.}~\bibnamefont{Wang}},
  \bibinfo{author}{\bibfnamefont{Y.}~\bibnamefont{Chong}},
  \bibinfo{author}{\bibfnamefont{J.~D.} \bibnamefont{Joannopoulos}},
  \bibnamefont{and}
  \bibinfo{author}{\bibfnamefont{M.}~\bibnamefont{Solja{\v{c}}i{\'{c}}}},
  \bibinfo{journal}{Nature} \textbf{\bibinfo{volume}{461}},
  \bibinfo{pages}{772} (\bibinfo{year}{2009}).

\bibitem[{\citenamefont{Brion et~al.}(1972)\citenamefont{Brion, Wallis,
  Hartstein, and Burstein}}]{Brion_1972}
\bibinfo{author}{\bibfnamefont{J.~J.} \bibnamefont{Brion}},
  \bibinfo{author}{\bibfnamefont{R.~F.} \bibnamefont{Wallis}},
  \bibinfo{author}{\bibfnamefont{A.}~\bibnamefont{Hartstein}},
  \bibnamefont{and} \bibinfo{author}{\bibfnamefont{E.}~\bibnamefont{Burstein}},
  \bibinfo{journal}{Physical Review Letters} \textbf{\bibinfo{volume}{28}},
  \bibinfo{pages}{1455} (\bibinfo{year}{1972}).

\bibitem[{\citenamefont{Wallis et~al.}(1974)\citenamefont{Wallis, Brion,
  Burstein, and Hartstein}}]{Wallis:Th}
\bibinfo{author}{\bibfnamefont{R.}~\bibnamefont{Wallis}},
  \bibinfo{author}{\bibfnamefont{J.}~\bibnamefont{Brion}},
  \bibinfo{author}{\bibfnamefont{E.}~\bibnamefont{Burstein}}, \bibnamefont{and}
  \bibinfo{author}{\bibfnamefont{A.}~\bibnamefont{Hartstein}},
  \textbf{\bibinfo{volume}{9}}, \bibinfo{pages}{3424} (\bibinfo{year}{1974}).

\bibitem[{\citenamefont{Tsakmakidis et~al.}(2017)\citenamefont{Tsakmakidis,
  Shen, Schulz, Zheng, Upham, Deng, Altug, Vakakis, and
  Boyd}}]{Tsakmakidis_2017}
\bibinfo{author}{\bibfnamefont{K.~L.} \bibnamefont{Tsakmakidis}},
  \bibinfo{author}{\bibfnamefont{L.}~\bibnamefont{Shen}},
  \bibinfo{author}{\bibfnamefont{S.~A.} \bibnamefont{Schulz}},
  \bibinfo{author}{\bibfnamefont{X.}~\bibnamefont{Zheng}},
  \bibinfo{author}{\bibfnamefont{J.}~\bibnamefont{Upham}},
  \bibinfo{author}{\bibfnamefont{X.}~\bibnamefont{Deng}},
  \bibinfo{author}{\bibfnamefont{H.}~\bibnamefont{Altug}},
  \bibinfo{author}{\bibfnamefont{A.~F.} \bibnamefont{Vakakis}},
  \bibnamefont{and} \bibinfo{author}{\bibfnamefont{R.~W.} \bibnamefont{Boyd}},
  \bibinfo{journal}{Science} \textbf{\bibinfo{volume}{356}},
  \bibinfo{pages}{1260} (\bibinfo{year}{2017}).

\bibitem[{\citenamefont{Xu et~al.}(2019)\citenamefont{Xu, Xiao, Wu, Zhang,
  Deng, and Shen}}]{Xu:Br}
\bibinfo{author}{\bibfnamefont{J.}~\bibnamefont{Xu}},
  \bibinfo{author}{\bibfnamefont{S.}~\bibnamefont{Xiao}},
  \bibinfo{author}{\bibfnamefont{C.}~\bibnamefont{Wu}},
  \bibinfo{author}{\bibfnamefont{H.}~\bibnamefont{Zhang}},
  \bibinfo{author}{\bibfnamefont{X.}~\bibnamefont{Deng}}, \bibnamefont{and}
  \bibinfo{author}{\bibfnamefont{L.}~\bibnamefont{Shen}},
  \textbf{\bibinfo{volume}{27}}, \bibinfo{pages}{10659} (\bibinfo{year}{2019}),
  ISSN \bibinfo{issn}{1094-4087}.

\bibitem[{\citenamefont{Xu et~al.}(2021)\citenamefont{Xu, He, Feng, Yong, Hong,
  Shen, and Zhou}}]{Xu:Sl}
\bibinfo{author}{\bibfnamefont{J.}~\bibnamefont{Xu}},
  \bibinfo{author}{\bibfnamefont{P.}~\bibnamefont{He}},
  \bibinfo{author}{\bibfnamefont{D.}~\bibnamefont{Feng}},
  \bibinfo{author}{\bibfnamefont{K.}~\bibnamefont{Yong}},
  \bibinfo{author}{\bibfnamefont{L.}~\bibnamefont{Hong}},
  \bibinfo{author}{\bibfnamefont{Y.}~\bibnamefont{Shen}}, \bibnamefont{and}
  \bibinfo{author}{\bibfnamefont{Y.}~\bibnamefont{Zhou}},
  \textbf{\bibinfo{volume}{29}}, \bibinfo{pages}{11328} (\bibinfo{year}{2021}),
  ISSN \bibinfo{issn}{1094-4087}.

\bibitem[{\citenamefont{Atkin et~al.}(2012)\citenamefont{Atkin, Berweger,
  Jones, and Raschke}}]{Atkin:Na}
\bibinfo{author}{\bibfnamefont{J.~M.} \bibnamefont{Atkin}},
  \bibinfo{author}{\bibfnamefont{S.}~\bibnamefont{Berweger}},
  \bibinfo{author}{\bibfnamefont{A.~C.} \bibnamefont{Jones}}, \bibnamefont{and}
  \bibinfo{author}{\bibfnamefont{M.~B.} \bibnamefont{Raschke}},
  \textbf{\bibinfo{volume}{61}}, \bibinfo{pages}{745} (\bibinfo{year}{2012}),
  ISSN \bibinfo{issn}{0001-8732, 1460-6976}.

\bibitem[{\citenamefont{Berweger et~al.}(2015)\citenamefont{Berweger, Weber,
  John, Velazquez, Pieterick, Sanford, Davydov, Brunschwig, Lewis, Wallis
  et~al.}}]{Berweger:Mi}
\bibinfo{author}{\bibfnamefont{S.}~\bibnamefont{Berweger}},
  \bibinfo{author}{\bibfnamefont{J.~C.} \bibnamefont{Weber}},
  \bibinfo{author}{\bibfnamefont{J.}~\bibnamefont{John}},
  \bibinfo{author}{\bibfnamefont{J.~M.} \bibnamefont{Velazquez}},
  \bibinfo{author}{\bibfnamefont{A.}~\bibnamefont{Pieterick}},
  \bibinfo{author}{\bibfnamefont{N.~A.} \bibnamefont{Sanford}},
  \bibinfo{author}{\bibfnamefont{A.~V.} \bibnamefont{Davydov}},
  \bibinfo{author}{\bibfnamefont{B.}~\bibnamefont{Brunschwig}},
  \bibinfo{author}{\bibfnamefont{N.~S.} \bibnamefont{Lewis}},
  \bibinfo{author}{\bibfnamefont{T.~M.} \bibnamefont{Wallis}},
  \bibnamefont{et~al.}, \textbf{\bibinfo{volume}{15}}, \bibinfo{pages}{1122}
  (\bibinfo{year}{2015}), ISSN \bibinfo{issn}{1530-6984, 1530-6992}.

\bibitem[{\citenamefont{Machida et~al.}(1987)\citenamefont{Machida, Yamamoto,
  and Itaya}}]{Machida:Ob}
\bibinfo{author}{\bibfnamefont{S.}~\bibnamefont{Machida}},
  \bibinfo{author}{\bibfnamefont{Y.}~\bibnamefont{Yamamoto}}, \bibnamefont{and}
  \bibinfo{author}{\bibfnamefont{Y.}~\bibnamefont{Itaya}},
  \bibinfo{journal}{Physical Review Letters} \textbf{\bibinfo{volume}{58}},
  \bibinfo{pages}{1000} (\bibinfo{year}{1987}).

\bibitem[{\citenamefont{Frascella et~al.}(2021)\citenamefont{Frascella, Agne,
  Khalili, and Chekhova}}]{Frascella:Ov}
\bibinfo{author}{\bibfnamefont{G.}~\bibnamefont{Frascella}},
  \bibinfo{author}{\bibfnamefont{S.}~\bibnamefont{Agne}},
  \bibinfo{author}{\bibfnamefont{F.~Y.} \bibnamefont{Khalili}},
  \bibnamefont{and} \bibinfo{author}{\bibfnamefont{M.~V.}
  \bibnamefont{Chekhova}}, \textbf{\bibinfo{volume}{7}}, \bibinfo{pages}{72}
  (\bibinfo{year}{2021}), ISSN \bibinfo{issn}{2056-6387}.

\bibitem[{\citenamefont{Tucker and Millea}(1978)}]{Tucker:Ph}
\bibinfo{author}{\bibfnamefont{J.~R.} \bibnamefont{Tucker}} \bibnamefont{and}
  \bibinfo{author}{\bibfnamefont{M.~F.} \bibnamefont{Millea}},
  \textbf{\bibinfo{volume}{33}}, \bibinfo{pages}{611} (\bibinfo{year}{1978}),
  ISSN \bibinfo{issn}{0003-6951, 1077-3118}.

\bibitem[{\citenamefont{Thapliyal et~al.}(2014)\citenamefont{Thapliyal, Pathak,
  Sen, and Pe{\v{r}}ina}}]{Thapliyal:Hi}
\bibinfo{author}{\bibfnamefont{K.}~\bibnamefont{Thapliyal}},
  \bibinfo{author}{\bibfnamefont{A.}~\bibnamefont{Pathak}},
  \bibinfo{author}{\bibfnamefont{B.}~\bibnamefont{Sen}}, \bibnamefont{and}
  \bibinfo{author}{\bibfnamefont{J.}~\bibnamefont{Pe{\v{r}}ina}},
  \bibinfo{journal}{Physical Review A} \textbf{\bibinfo{volume}{90}},
  \bibinfo{pages}{013808} (\bibinfo{year}{2014}).

\bibitem[{\citenamefont{Silveirinha and Engheta}(2006)}]{silveirinha_2006}
\bibinfo{author}{\bibfnamefont{M.}~\bibnamefont{Silveirinha}} \bibnamefont{and}
  \bibinfo{author}{\bibfnamefont{N.}~\bibnamefont{Engheta}},
  \bibinfo{journal}{Physical Review Letters} \textbf{\bibinfo{volume}{97}},
  \bibinfo{pages}{157403} (\bibinfo{year}{2006}).

\bibitem[{\citenamefont{Liu et~al.}(2008)\citenamefont{Liu, Cheng, Hand, Mock,
  Cui, Cummer, and Smith}}]{liu_2008}
\bibinfo{author}{\bibfnamefont{R.}~\bibnamefont{Liu}},
  \bibinfo{author}{\bibfnamefont{Q.}~\bibnamefont{Cheng}},
  \bibinfo{author}{\bibfnamefont{T.}~\bibnamefont{Hand}},
  \bibinfo{author}{\bibfnamefont{J.~J.} \bibnamefont{Mock}},
  \bibinfo{author}{\bibfnamefont{T.~J.} \bibnamefont{Cui}},
  \bibinfo{author}{\bibfnamefont{S.~A.} \bibnamefont{Cummer}},
  \bibnamefont{and} \bibinfo{author}{\bibfnamefont{D.~R.} \bibnamefont{Smith}},
  \bibinfo{journal}{Physical Review Letters} \textbf{\bibinfo{volume}{100}},
  \bibinfo{pages}{023903} (\bibinfo{year}{2008}).

\bibitem[{\citenamefont{Liu et~al.}(2009)\citenamefont{Liu, Hu, Zhao, and
  Luo}}]{liu_2009}
\bibinfo{author}{\bibfnamefont{L.}~\bibnamefont{Liu}},
  \bibinfo{author}{\bibfnamefont{C.}~\bibnamefont{Hu}},
  \bibinfo{author}{\bibfnamefont{Z.}~\bibnamefont{Zhao}}, \bibnamefont{and}
  \bibinfo{author}{\bibfnamefont{X.}~\bibnamefont{Luo}},
  \bibinfo{journal}{Optics Express} \textbf{\bibinfo{volume}{17}},
  \bibinfo{pages}{12183} (\bibinfo{year}{2009}).

\bibitem[{\citenamefont{Jingjing et~al.}(2010)\citenamefont{Jingjing, Yu, and
  Niels~Asger}}]{Jing_2010}
\bibinfo{author}{\bibfnamefont{Z.}~\bibnamefont{Jingjing}},
  \bibinfo{author}{\bibfnamefont{L.}~\bibnamefont{Yu}}, \bibnamefont{and}
  \bibinfo{author}{\bibfnamefont{M.}~\bibnamefont{Niels~Asger}},
  \bibinfo{journal}{Optics Express} \textbf{\bibinfo{volume}{18}},
  \bibinfo{pages}{3864} (\bibinfo{year}{2010}).

\bibitem[{\citenamefont{Li et~al.}(2017)\citenamefont{Li, Sun, Sun, Wang, Song,
  Liu, Chen, and Gu}}]{li_2017}
\bibinfo{author}{\bibfnamefont{Z.}~\bibnamefont{Li}},
  \bibinfo{author}{\bibfnamefont{Y.}~\bibnamefont{Sun}},
  \bibinfo{author}{\bibfnamefont{H.}~\bibnamefont{Sun}},
  \bibinfo{author}{\bibfnamefont{K.}~\bibnamefont{Wang}},
  \bibinfo{author}{\bibfnamefont{J.}~\bibnamefont{Song}},
  \bibinfo{author}{\bibfnamefont{L.}~\bibnamefont{Liu}},
  \bibinfo{author}{\bibfnamefont{X.}~\bibnamefont{Chen}}, \bibnamefont{and}
  \bibinfo{author}{\bibfnamefont{C.}~\bibnamefont{Gu}},
  \bibinfo{journal}{Journal of Physics D: Applied Physics}
  \textbf{\bibinfo{volume}{50}}, \bibinfo{pages}{375105}
  (\bibinfo{year}{2017}).

\bibitem[{\citenamefont{Xu et~al.}(2018)\citenamefont{Xu, Liu, Li, Zhao, Liu,
  and Yin}}]{xu_2018}
\bibinfo{author}{\bibfnamefont{Z.}~\bibnamefont{Xu}},
  \bibinfo{author}{\bibfnamefont{S.}~\bibnamefont{Liu}},
  \bibinfo{author}{\bibfnamefont{S.}~\bibnamefont{Li}},
  \bibinfo{author}{\bibfnamefont{H.}~\bibnamefont{Zhao}},
  \bibinfo{author}{\bibfnamefont{L.}~\bibnamefont{Liu}}, \bibnamefont{and}
  \bibinfo{author}{\bibfnamefont{X.}~\bibnamefont{Yin}},
  \bibinfo{journal}{Applied Physics Express} \textbf{\bibinfo{volume}{11}},
  \bibinfo{pages}{042002} (\bibinfo{year}{2018}).

\bibitem[{\citenamefont{Liang et~al.}(2015)\citenamefont{Liang, Wang, Zhao,
  Wang, Yao, Gao, Luo, Gao, Zhao, and Luo}}]{Liang:Sq}
\bibinfo{author}{\bibfnamefont{G.}~\bibnamefont{Liang}},
  \bibinfo{author}{\bibfnamefont{C.}~\bibnamefont{Wang}},
  \bibinfo{author}{\bibfnamefont{Z.}~\bibnamefont{Zhao}},
  \bibinfo{author}{\bibfnamefont{Y.}~\bibnamefont{Wang}},
  \bibinfo{author}{\bibfnamefont{N.}~\bibnamefont{Yao}},
  \bibinfo{author}{\bibfnamefont{P.}~\bibnamefont{Gao}},
  \bibinfo{author}{\bibfnamefont{Y.}~\bibnamefont{Luo}},
  \bibinfo{author}{\bibfnamefont{G.}~\bibnamefont{Gao}},
  \bibinfo{author}{\bibfnamefont{Q.}~\bibnamefont{Zhao}}, \bibnamefont{and}
  \bibinfo{author}{\bibfnamefont{X.}~\bibnamefont{Luo}},
  \bibinfo{journal}{Advanced Optical Materials} \textbf{\bibinfo{volume}{3}},
  \bibinfo{pages}{1248} (\bibinfo{year}{2015}).

\bibitem[{\citenamefont{Iranzo et~al.}(2018)\citenamefont{Iranzo, Nanot, Dias,
  Epstein, Peng, Efetov, Lundeberg, Parret, Osmond, Hong et~al.}}]{iranzo_2018}
\bibinfo{author}{\bibfnamefont{D.~A.} \bibnamefont{Iranzo}},
  \bibinfo{author}{\bibfnamefont{S.}~\bibnamefont{Nanot}},
  \bibinfo{author}{\bibfnamefont{E.~J.} \bibnamefont{Dias}},
  \bibinfo{author}{\bibfnamefont{I.}~\bibnamefont{Epstein}},
  \bibinfo{author}{\bibfnamefont{C.}~\bibnamefont{Peng}},
  \bibinfo{author}{\bibfnamefont{D.~K.} \bibnamefont{Efetov}},
  \bibinfo{author}{\bibfnamefont{M.~B.} \bibnamefont{Lundeberg}},
  \bibinfo{author}{\bibfnamefont{R.}~\bibnamefont{Parret}},
  \bibinfo{author}{\bibfnamefont{J.}~\bibnamefont{Osmond}},
  \bibinfo{author}{\bibfnamefont{J.-Y.} \bibnamefont{Hong}},
  \bibnamefont{et~al.}, \bibinfo{journal}{Science}
  \textbf{\bibinfo{volume}{360}}, \bibinfo{pages}{291} (\bibinfo{year}{2018}).

\bibitem[{\citenamefont{Deng et~al.}(2015)\citenamefont{Deng, Hong, Zheng, and
  Shen}}]{Deng_2015}
\bibinfo{author}{\bibfnamefont{X.}~\bibnamefont{Deng}},
  \bibinfo{author}{\bibfnamefont{L.}~\bibnamefont{Hong}},
  \bibinfo{author}{\bibfnamefont{X.}~\bibnamefont{Zheng}}, \bibnamefont{and}
  \bibinfo{author}{\bibfnamefont{L.}~\bibnamefont{Shen}},
  \bibinfo{journal}{Applied Optics} \textbf{\bibinfo{volume}{54}},
  \bibinfo{pages}{4608} (\bibinfo{year}{2015}).

\bibitem[{\citenamefont{Shen et~al.}(2019)\citenamefont{Shen, Shen, Min, Xu,
  Wu, Deng, and Xiao}}]{Shen_2019}
\bibinfo{author}{\bibfnamefont{Q.}~\bibnamefont{Shen}},
  \bibinfo{author}{\bibfnamefont{L.}~\bibnamefont{Shen}},
  \bibinfo{author}{\bibfnamefont{W.}~\bibnamefont{Min}},
  \bibinfo{author}{\bibfnamefont{J.}~\bibnamefont{Xu}},
  \bibinfo{author}{\bibfnamefont{C.}~\bibnamefont{Wu}},
  \bibinfo{author}{\bibfnamefont{X.}~\bibnamefont{Deng}}, \bibnamefont{and}
  \bibinfo{author}{\bibfnamefont{S.}~\bibnamefont{Xiao}},
  \bibinfo{journal}{Optical Materials Express} \textbf{\bibinfo{volume}{9}},
  \bibinfo{pages}{4399} (\bibinfo{year}{2019}).

\bibitem[{\citenamefont{Xu et~al.}(2020)\citenamefont{Xu, Deng, Zhang, Wu,
  Wubs, Xiao, and Shen}}]{Xu:Ul}
\bibinfo{author}{\bibfnamefont{J.}~\bibnamefont{Xu}},
  \bibinfo{author}{\bibfnamefont{X.}~\bibnamefont{Deng}},
  \bibinfo{author}{\bibfnamefont{H.}~\bibnamefont{Zhang}},
  \bibinfo{author}{\bibfnamefont{C.}~\bibnamefont{Wu}},
  \bibinfo{author}{\bibfnamefont{M.}~\bibnamefont{Wubs}},
  \bibinfo{author}{\bibfnamefont{S.}~\bibnamefont{Xiao}}, \bibnamefont{and}
  \bibinfo{author}{\bibfnamefont{L.}~\bibnamefont{Shen}},
  \textbf{\bibinfo{volume}{22}}, \bibinfo{pages}{025003}
  (\bibinfo{year}{2020}), ISSN \bibinfo{issn}{2040-8978}.

\bibitem[{\citenamefont{Pozar}(2011)}]{Pozar:Mi}
\bibinfo{author}{\bibfnamefont{D.~M.} \bibnamefont{Pozar}},
  \emph{\bibinfo{title}{Microwave Engineering}} (\bibinfo{publisher}{{John
  wiley \& sons}}, \bibinfo{year}{2011}).

\end{thebibliography}

\end{document}